\documentclass[fleqn,twoside]{article}
\usepackage[headings]{espcrc2}

\readRCS
$Id: espcrc2.tex,v 1.2 2004/02/24 11:22:11 spepping Exp $
\usepackage{graphicx}
\usepackage[figuresright]{rotating}

\newcommand{\AmS}{{\protect\the\textfont2
  A\kern-.1667em\lower.5ex\hbox{M}\kern-.125emS}}

\title{Sensitivity to anomalous quartic gauge couplings 
in~photon-photon~interactions at~the~LHC
\newline {\small Contribution to the CERN workshop on 
"High energy photon collisions at the LHC", 22-25$^{th}$ April 2008}}

\author{
T. Pierzcha\l a\address[UCL]{Universit\'e catholique de Louvain,
    Center for Particle Physics and Phenomenology (CP3)\\
    Chemin du Cyclotron 2, 1348 Louvain-la-Neuve, Belgium}\thanks{Email: Tomasz.Pierzchala@uclouvain.be}
    and K. Piotrzkowski\addressmark[UCL]
}
                       
\runtitle{Anomalous quartic gauge couplings 
in~$\gamma\gamma$~interactions at~the~LHC}
\runauthor{T. Pierzcha\l a and K. Piotrzkowski}

\begin{document}

\begin{abstract}
The exclusive two-photon production at the LHC of pairs of W and Z bosons provides a novel and unique
test-ground for the electroweak gauge boson sector. In particular it offers, thanks to high
$\gamma\gamma$ center-of-mass energies, large and direct sensitivity to the anomalous quartic gauge
couplings otherwise very difficult to investigate at the LHC. An initial analysis has been performed assuming 
leptonic decays and generic acceptance cuts. Simulation of a simple counting experiment has shown 
for the integrated luminosity of 10 fb$^{-1}$ at least four thousand times larger sensitivity to the genuine 
quartic couplings, $a_0^W$,  $a_0^Z$, $a_C^W$ and $a_C^Z$, than those obtained at LEP. The impact of the unitarity constraints on 
the estimated limits has been studied using the dipole form-factors. Finally, differential distributions 
of the decay leptons have been provided to illustrate the potential for further improvements of the 
sensitivities.
\vspace{1pc}
\end{abstract}

\maketitle
\renewcommand{\arraystretch}{1.3}
\section{LHC as a photon collider}
The $\gamma\gamma$ exclusive production of pairs of charged particles offers 
interesting potential for signals of new physics at the LHC. In a recent paper 
\cite{paper}, the initial comprehensive studies of high energy photon interactions at the LHC were 
reported. In the present contribution, the selected results on the gauge boson pairs discussed in Ref. 
\cite{paper} are introduced and supplemented by new results.

The exclusive two-photon production, 
$pp\rightarrow pXp$, provides clean experimental conditions, thanks to absence of the proton remnants. 
Well defined final states can be then selected, and precisely reconstructed. 
Moreover, detection of the two final state protons, scattered at almost zero-degree angle, 
in the dedicated very forward detectors (VFDs), provides another striking signature, effective also at high
luminosity and with large event pile-up \cite{piotr,fp420}. In addition, the photon energies can be then measured and
used for the event kinematics reconstruction. Finally, virtualities of the exchanged photons are on average very small,
and are limited from above due to the proton electromagnetic form-factors, allowing for treating the LHC protons as sources of
quasi-real photons.

\begin{figure}[htb]
\vspace{9pt}
  \includegraphics[width=19pc]{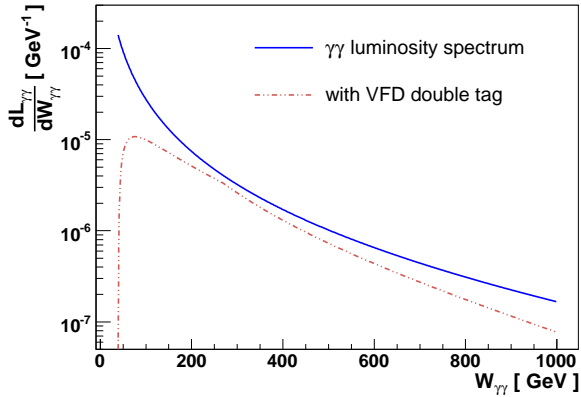}
\caption{Elastic luminosity spectrum of photon-photon collisions at the LHC
assuming the maximal photon virtuality $Q_{max}^2=2~\textrm{GeV}^2$ (solid line).
The luminosity spectrum assuming the photon tagging range 
$20~\textrm{GeV}~<~E_{\gamma}~<~900~\textrm{GeV}$ is also shown (dashed  line).}
\label{fig:DiffLumi}
\end{figure}

The cross sections of two-photon pair production are in general determined by the mass, spin and charge of 
the produced particles, so the rate of produced particles at the LHC can be well predicted 
using the Equivalent Photon Approximation for the equivalent photon fluxes \cite{budnev}. 
For proton-proton collisions at $\sqrt{s}=14$~TeV the EPA 
predicts the photon-photon luminosity spectrum as shown in Fig.~\ref{fig:DiffLumi}, where
$d L_{\gamma\gamma}/dW_{\gamma\gamma}$ is defined by the relation between the proton-proton
and $\gamma\gamma$ cross-sections assuming a minimal center-of-mass energy $W_0$:
$$
\sigma_{pp}=\int_{W_0}^{\sqrt{s}}\sigma_{\gamma\gamma}~
\frac{d L_{\gamma\gamma}}{dW_{\gamma\gamma}}~dW_{\gamma\gamma}~.
$$

\begin{figure}[htb]
\vspace{9pt}
  \includegraphics[width=19pc]{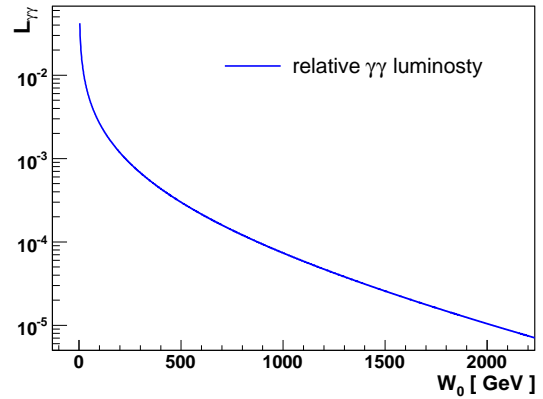}
\caption{Relative elastic $\gamma\gamma$ luminosity for photon collisions at the center-of-mass energy above $W_0$, obtained  
using the EPA for proton collisions at $\sqrt{s}=14~TeV$, and assuming the maximal photon virtuality 
$Q_{max}^2=2~\textrm{GeV}^2$.}
\label{fig:RelativeLumi}
\end{figure}

A set of VFDs at 220~m or 420~m from the LHC interaction points will be capable of tagging photon interactions 
within the wide photon energy range of $20~\textrm{GeV}<E_{\gamma}<900~\textrm{GeV}$ \cite{hector,xavier}.
In Fig.~\ref{fig:DiffLumi} also the $\gamma\gamma$ luminosity spectrum is shown assuming double tagging 
(i.e. requesting both forward protons to be detected). One should note that apart from such $elastic$
two-photon processes where both protons have survived the interaction, the $inelastic$ production can
also be considered, when at least one of the two protons dissociates into a low mass state. 
The corresponding two-photon luminosity increases then by about a factor of three \cite{piotr}. 
By integrating the luminosity spectrum above some minimal center-of-mass energy $W_0$, one
can introduce the relative photon-photon luminosity $L_{\gamma\gamma}$, 
shown in Fig.~\ref{fig:RelativeLumi}. Effectively, $L_{\gamma\gamma}$ gives a fraction of the proton-proton
luminosity which is available for $\gamma\gamma$ collisions, and is especially useful if a given photon-photon
cross-section is approximately constant as a function of $W_{\gamma\gamma}$. For example, the relative photon-photon 
luminosity at the LHC is equal to 1\% for $W_0=23$~GeV (i.e. for $W_{\gamma\gamma}>23$~GeV), and 0.1\% for $W_0=225$~GeV.
Given the very large LHC luminosity, this leads to significant event rates of high-energy processes 
with relatively small photon-photon cross sections. In the following only the proton cross-sections are quoted.

One needs however to consider corrections beyond the EPA due to a possibility of strong interactions between protons, 
or the so-called rescattering effects. The resulting suppression of the cross sections weakly depends on 
the invariant mass of the exclusively produced state $X$, and for the processes discussed in the following, 
as $pp\rightarrow pWWp$, it is estimated to be about 15\% \cite{khoze}. This correction is ignored in the present analysis.
In addition, one should stress that potentially dangerous background due to the exclusive diffractive production
is heavily suppressed for non strongly-interacting particles and can be safely neglected. 
For example, the gluon mediated exclusive production of W boson pairs, 
is about 100 times smaller than the two-photon production at the LHC \cite{khoze2}.

Anomalous quartic couplings of the electroweak gauge bosons could directly reveal 
the exchange of new heavy bosons \cite{Belanger:1992qh,wudka}, and the two-photon exclusive production of WW 
and ZZ pairs is particularly well suited for studies of the quartic
couplings $\gamma\gamma WW$ and $\gamma\gamma ZZ$. In the Standard Model (SM), the cross section of the exclusive 
two-photon production $pp \rightarrow pWWp$ 
is large about 108~fb$^{-1}$ \cite{paper}. This means that unique, high statistics tests of the gauge boson sector in the SM
can be performed at the LHC. In contrast, the cross section of the exclusive two-photon production $pp \rightarrow pZZp$
is very small in the SM, since this proccess is not allowed at the tree level. It means that an observation of even a few 
events of this type could signal the new physics. 

\section{Anomalous quartic gauge couplings}

The anomalous quartic gauge couplings (AQGCs) can be introduced in various ways, usually by building an effective lagrangian which models a low energy 
behavior of a wide class of possible extensions of the SM. For example, one can introduce new terms in such a lagrangian, which are allowed by 
the local $\mathrm{U}(1)_Y\times\mathrm{SU}(2)_L$ gauge invariance, and then consider two scenarios, with \cite{andreas} and without \cite{oscar} the Higgs boson.
In the present analysis, the  phenomenological lagrangians are used, which allow for genuine anomalous quartic vector boson couplings,
without need for associated trilinear gauge couplings. First, the simplest lagrangian term of power six in energy for two photon 
interaction with weak bosons has to conserve local $\mathrm{U}(1)_{em}$ and custodial SU(2)$_c$. Then, by imposing conservation of discrete C and P 
symmetries one finally obtains two new terms \cite{Belanger:1992qh,Eboli:2000ad}:
\[
L^0_6 = -\frac{e^2}{16}\frac{a_0}{\Lambda^2}F_{\mu\nu}F^{\mu\nu}
\vec{W}^\alpha \cdot \vec{W}_\alpha
\]
\begin{equation}
L^C_6 = -\frac{e^2}{16}\frac{a_C}{\Lambda^2}F_{\mu\alpha}F^{\mu\beta}
\vec{W}^\alpha \cdot \vec{W}_\beta,
\end{equation}
where $e$ is the electron charge and $\Lambda$ is the energy scale of the new physics. When the global SU(2)$_c$ symmetry is
not imposed then one can distinguish the neutral and charged couplings of the W and Z bosons. Since this more general model has been 
assumed for studies at LEP \cite{OPAL.limits}, to allow for a direct comparison of sensitivities, this is also assumed in the following:
\begin{eqnarray}
\mathcal{L}_6^0 &~~=~~&  - \frac{e^2}{8} {\bf\frac{a_0^{\mathrm{W}}}{\Lambda^2}} F_{\mu\nu}F^{\mu\nu} {W}^{+\alpha}{W}^-_{\alpha}\nonumber\\
& & - \frac{e^2}{16\cos^2\theta_W} {\bf\frac{a_0^{\mathrm{Z}}}{\Lambda^2}} F_{\mu\nu}F^{\mu\nu} {Z}^{\alpha}{Z}_{\alpha},
 \nonumber \\
\mathcal{L}_6^{\mathrm{c}} &~~=~~&  - \frac{e^2}{16} {\bf\frac{a_{\mathrm{c}}^{\mathrm{W}}}{\Lambda^2}} F_{\mu\alpha}F^{\mu\beta} ({W}^{+\alpha}{W}^-_{\beta}+
W^{-\alpha}W^+_{\beta}) \nonumber\\
& & - \frac{e^2}{16\cos^2\theta_W} {\bf\frac{a_{\mathrm{c}}^{\mathrm{Z}}}{\Lambda^2}} F_{\mu\alpha}F^{\mu\beta} {Z}^{\alpha}{Z}_{\beta}.
\label{eq.AnomalLagrangian}
\end{eqnarray}
Using this formalism, one obtains a general relation for the $WW$ and $ZZ$ cross sections (which is also valid after applying the acceptance cuts) 
as a function of the anomalous parameters:
\begin{equation}
\sigma = \sigma_{SM} + \sigma_0 a_0 + \sigma_{00} a^2_0 + \sigma_{c} a_c
+ \sigma_{cc} a^2_c + \sigma_{0c} a_0 a_c ~~~\\
\label{ex.anom}
\end{equation}
which corresponds, for a fixed cross section $\sigma$, to an ellipse on the $a_0$, $a_c$ plane. 
\section{Estimation of sensitivity}
Simulation of the exclusive two-photon pair production at the LHC was performed using the modified 
MadGraph/MadEvent~\cite{mad} and CalcHep \cite{calc} packages. The generated events were then passed to
the modified Pythia generator \cite{pythia} to allow for decays and hadronisation.
The sensitivity to the anomalous quartic vector boson couplings at the LHC has been investigated using the signature
of two opposite charge leptons ($e$ or $\mu$) within the generic lepton acceptance window -- $|\eta|<2.5$ and $p_T>10$~GeV.
In the $WW$ case it corresponds to the subprocesses $\gamma\gamma\rightarrow W^+W^-\rightarrow l^+l^-\nu\bar{\nu}$ 
while in the $ZZ$ case to the subprocesses $\gamma\gamma\rightarrow ZZ\rightarrow l^+l^- j j$.  
It is assumed that both processes are background free.
Under this condition, the upper limits of number of events $\lambda^{up}$ at the 95\% confidence level (CL) 
were calculated assuming the number of observed events equal to the $SM$ prediction 
$\mathrm{N_{obs} = \sigma^{\textsc{sm}}_{acc}~L}$ for a given integrated luminosity L. 
\begin{equation}
\mathrm{\sum^{N_{obs}}_{k=0} P_{Poisson}(\lambda^{up} = 
\sigma ^{up}~L; k) = 1 - CL}
\label{eq.simple-Poisson}
\end{equation}
This is a simplified approach and for more precise analysis one should use formula from \cite{Zech:1988un}, 
however the difference in the obtained limits is small, less than 20\%. 

The expected 95$\%$ CL limits $\mathrm{\lambda^{up}}$ have been then used to calculate  
the upper limits on the observed cross section $\mathrm{\sigma^{up}}$ at the 95\% CL
for the integrated luminosity L~=~1~fb$^{-1}$ and L~=~10~fb$^{-1}$, shown in the Table~\ref{tab.xs-limits}.

\begin{figure}[htb]
\vspace{9pt}
  \includegraphics[width=19pc]{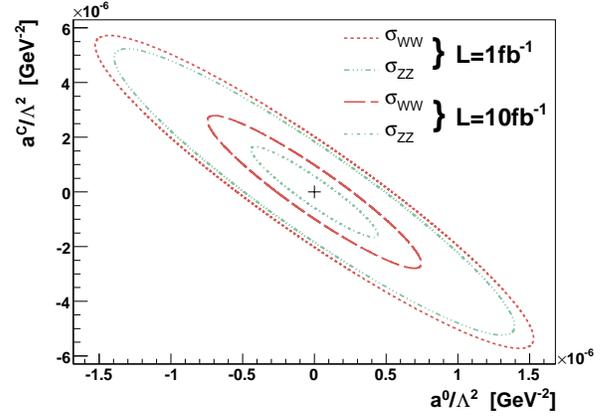}
\caption{Profiles of the 95$\%$ CL upper limits of the cross section 
after the acceptance cuts for 
$pp (\gamma\gamma\rightarrow W^+W^-\rightarrow$ $l^+l^-\nu\bar{\nu}) pp$  
and $pp (\gamma\gamma\rightarrow ZZ\rightarrow$ $l^+l^- j j) pp$ 
as a function of relevant anomalous couplings  $\mathrm{a_0^W/ \Lambda^2}$, 
$\mathrm{a_{\mathrm{c}}^W/ \Lambda^2}$, $\mathrm{a_0^Z/ \Lambda^2}$ and 
$\mathrm{a_{\mathrm{c}}^Z/ \Lambda^2}$. 
The contours are shown assuming two values of the integrated $pp$ luminosity.}
\label{fig:anomal_2D}
\end{figure}

\begin{figure}[htb]
\vspace{9pt}
  \includegraphics[width=19pc]{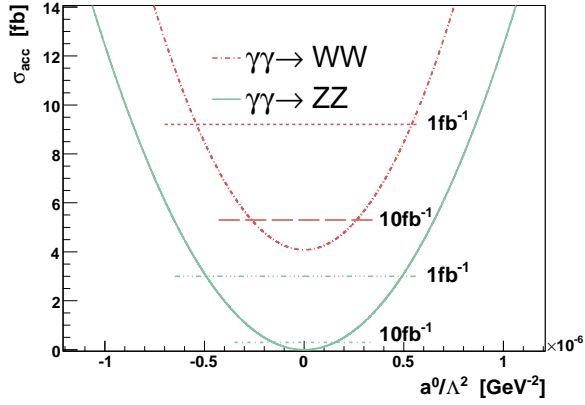}
\caption{Cross sections of $pp (\gamma\gamma\rightarrow W^+W^-\rightarrow$ $l^+l^-\nu\bar{\nu}) pp$ 
and $pp (\gamma\gamma\rightarrow ZZ\rightarrow$ $l^+l^- j j) pp$ 
after the acceptance cuts are shown
as a function of the genuine anomalous quartic vector boson couplings
$\mathrm{a_0^W / \Lambda^2}$ and $\mathrm{a_0^Z/ \Lambda^2}$ (for $\mathrm{a_C^W/ \Lambda^2}=\mathrm{a_C^Z/ \Lambda^2=0}$),
together with the upper cross section limits at CL=95$\%$ (horizontal lines). 
}
\label{fig:anomal_1D}
\end{figure}

\begin{table}[h!]
\caption{Expected 95$\%$ CL upper limits for the cross sections after acceptance cuts for chosen subprocesses
$\gamma\gamma\rightarrow W^+W^-\rightarrow l^+l^-\nu\bar{\nu}$ and $\gamma\gamma\rightarrow ZZ\rightarrow l^+l^- j j$. The assumed numbers
of the observed $WW$ events correspond to the $SM$ cross section calculated using \textsc{MG/ME} and the acceptance cuts.
}
\label{tab.xs-limits}
\begin{center}
 \begin{tabular}{ccc}
 \hline
 & $\gamma\gamma\rightarrow W^+W^-$ & $\gamma\gamma\rightarrow ZZ$ \\
 $\mathrm{\sigma^{up}~[fb]}$ & & $\mathrm{N_{obs}= 0}$ \\
 & $\mathrm{\sigma^{\textsc{sm}}_{acc}=4.081~fb}$ & $\mathrm{\lambda^{up} = 2.996}$  \\ \hline

      $\mathrm{L = 1~fb^{-1}}$    &  $\mathrm{9.2}$ & $\mathrm{3.0}$  \\
      $\mathrm{L = 10~fb^{-1}}$   &  $\mathrm{5.3}$ & $\mathrm{0.30}$ \\ 
         \hline
\end{tabular}
\end{center}
\end{table}
The calculated cross section upper limits can be directly converted to the limits of the anomalous quartic couplings
as presented at the Fig.~\ref{fig:anomal_2D} where the 95\% CL contours are shown.
In Fig.~\ref{fig:anomal_1D} one parameter limits (with the other anomalous coupling set to zero) are shown, and the obtained limits are
quoted in the Table~\ref{tab.1st-limits}.

\begin{table}[h!]
\begin{center}
\caption{Expected one-parameter limits for anomalous quartic vector boson couplings at 
95$\%$ CL for two values of the integrated luminosity.}
\label{tab.1st-limits}
\begin{tabular}{ccc}
\hline
Coupling & \multicolumn{2}{c}{ Limits $\mathrm{[10^{-6}~\textrm{GeV}^{-2}~]}$} \\  
         &$\mathrm{L = 1~fb^{-1}}$ & $ \mathrm{L = 10~fb^{-1}}$ \\
\hline
 $ \mathrm{|a^Z_0/\Lambda^2|} $& $ 0.49 $ & $ 0.16 $ \\
 $ \mathrm{|a^Z_C/\Lambda^2|} $& $ 1.84 $ & $ 0.58 $ \\
 $ \mathrm{|a^W_0/\Lambda^2|} $& $ 0.54 $ & $ 0.27 $ \\
 $ \mathrm{|a^W_C/\Lambda^2|} $& $ 2.02 $ & $ 0.99 $ \\
\hline
\end{tabular}
\end{center}
\end{table}
The obtained limits are about 40~000 times better than the best limits established at LEP2
\cite{OPAL.limits} clearly showing large and unique potential of such studies at the LHC.
\section{Impact of unitarity bound condition}
The lagrangian terms in Eq.(\ref{eq.AnomalLagrangian}), do not preserve the SU(2)$_L$ local symmetry. 
In consequence, for the center-of-mass energies $W_{\gamma\gamma}\gg 2 M_W$ the scattering amplitude
will grow and eventually will violate the unitarity condition. It is therefore necessary to investigate
the impact of the unitarity constraint on the derived limits. For simplicity, the unitarity violation is
checked only for the process $\gamma\gamma\rightarrow W^+ W^-$. The $\gamma\gamma \rightarrow 
\gamma\gamma$ and $\gamma\gamma \rightarrow ZZ$ (i.e. the anomalous coupling $\gamma\gamma ZZ$ is 
set to zero) processes are neglected, as well as the two-photon exclusive production of fermion pairs as it 
drops fast with $W_{\gamma\gamma}$ \cite{paper}.
For spin one particles one obtains the partial wave amplitudes $a_J$ using the Legendre polynomial $P_J$: 
\begin{equation}
a_{J}(\sqrt{s}) =\frac{1}{32\pi}\int_{-1}^{1}du~{\cal M}(\sqrt{s},u,a^W_0,a^W_C)P_{J}(u),~
\label{eq.a_J}
\end{equation}
where {\it u} is the cosine of the W boson polar angle in the $\gamma\gamma$ center-of-mass system,
and ${\cal M}$ is the amplitude of the process $\gamma\gamma\rightarrow W^+ W^-$.
Then, the unitarity bound condition has a form:
\begin{equation}
\beta \sum_{pol}|a_{J}(\sqrt{s},a^W_0,a^W_C)|^2 \le (1/2)^2
\label{eq.unitarity-bound}
\end{equation}
 where $\beta = \sqrt{1-4m_{W}^2/s}$ is the Lorentz velocity of a W boson 
in the cms frame and {\it pol} stands for polarization states in 
$\gamma\gamma\rightarrow W^+W^-$. 

\begin{figure}[htb]
\vspace{9pt}
  \includegraphics[width=19pc]{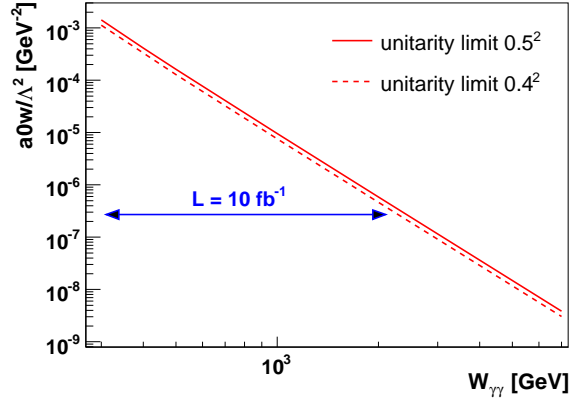}
  \caption{Unitarity limit for the W neutral anomalous coupling $a_0^W/\Lambda^2$ 
calculated according to Eq.~\ref{eq.unitarity-bound} for J=0 (full~line), and
assuming that the left hand side of Eq.~\ref{eq.unitarity-bound} is smaller than $(0.4)^2$
(dashed line). The arrow indicates the obtained limit at 95\% CL from Tab.~\ref{tab.1st-limits} for $10$~fb$^{-1}$. 
}
\label{fig:a0w_noFF}
\end{figure}
\begin{figure}[htb]
\vspace{9pt}
  \includegraphics[width=19pc]{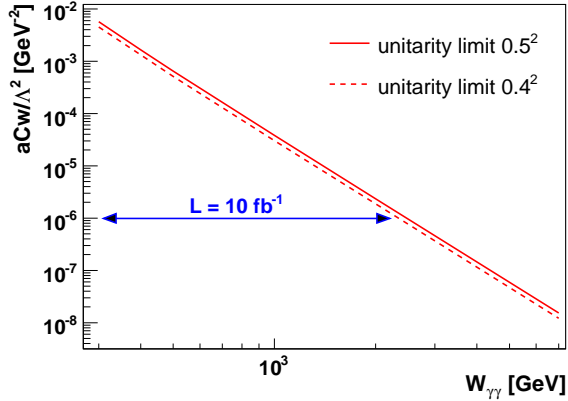}
\caption{Analogous plot to Fig.~\ref{fig:a0w_noFF} but for the W charged
anomalous coupling $a_C^W/\Lambda^2$. Also in this case, unitarity will
be violated for $W_{\gamma\gamma}$ above 2~TeV.}
\label{fig:aCw_noFF}
\end{figure}

Unitarity bounds of Eq.~\ref{eq.unitarity-bound} for J=0 as a function of $W_{\gamma\gamma}$
are shown for both neutral, in Fig.~\ref{fig:a0w_noFF}, and charged, 
in Fig.~\ref{fig:aCw_noFF}, AQGCs together with the obtained limits of anomalous couplings for the integrated
luminosity $L=10$~fb$^{-1}$. 
It shows that for these limits the unitarity will be violated when $W_{\gamma\gamma}$ is above 2~TeV, also
if one allows for a contribution of some other channels, by requesting the left hand side 
of Eq.~\ref{eq.unitarity-bound} to be smaller than $(0.4)^2$ instead of $(0.5)^2$.
Other $WW$ states with different total angular momentum do not violate the unitarity earlier, and 
$a_{1}$ is close to 0 when $a_{2}$ is much smaller than $a_{0}$. 

\begin{figure}[htb]
\vspace{9pt}
  \includegraphics[width=19pc]{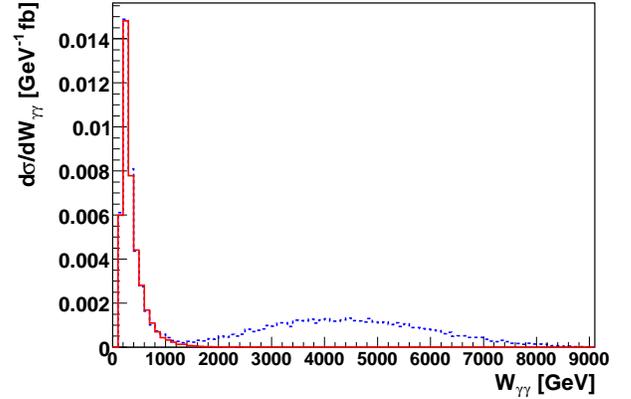}
\caption{Differential cross section $d\sigma/dW_{\gamma\gamma}$ for
$\gamma\gamma \rightarrow W^+W^- \rightarrow l^+l^-$ after acceptance cuts, 
for the SM (solid line) and for the AQGC case (dashed line), assuming $a_0^W/\Lambda^2=0.54\cdot 10^{-6}$~GeV$^{-2}$ 
and $a_C^W/\Lambda^2=0$.}
\label{fig.Diff_W}
\end{figure}

The unitarity violation occurs due to the increasing contribution of longitudinally 
polarized W bosons, which in the SM is fully compensated thanks to the local SU(2)$_L$ symmetry
and the trilinear gauge couplings. As a consequence, one has to verify if the
limits quoted in Tab.~\ref{tab.1st-limits} are not driven by $\gamma\gamma$ interactions at
$W_{\gamma\gamma}$ above 2~TeV. As one can see in Fig.~\ref{fig.Diff_W},
this is actually the case, since the anomalous cross-section is large 
for the energies well above 2 TeV. This can be avoided in two ways.
The double tagging using very forward detectors would limit
the maximal $\gamma\gamma$ center-of-mass energy to 1.8~TeV~\cite{xavier}. One should note that
the tagging efficiency in this energy range is high, about 80\% \cite{schul}.
In other case, one can introduce dipole form-factors for each anomalous coupling, as in Eq.~\ref{eq.formfactor}.
The form-factors suppress the anomalous amplitudes when the energies are close to the new physics
energy scale $\Lambda$:
\begin{equation}
{\bf a}\rightarrow \frac{\bf a}{\left(1+ W_{\gamma\gamma}^2/{\Lambda^2} \right)^X},
\label{eq.formfactor}
\end{equation}
where the power $X$ is chosen so to preserve the unitarity -- in the present analysis $X=2$.
\section{Expected limits for anomalous quartic gauge couplings}
\begin{figure}[htb]
\vspace{9pt}
  \includegraphics[width=19pc]{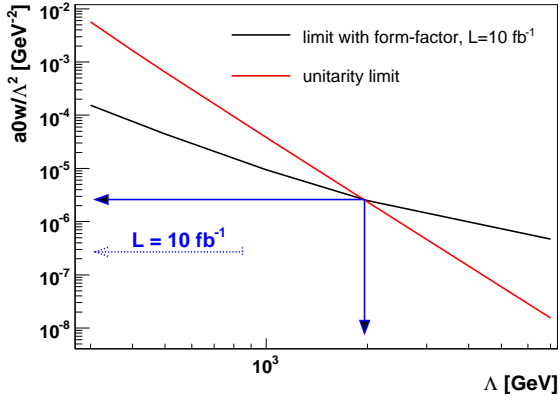}
\caption{The 95\%~CL limits of the W neutral anomalous quartic coupling 
(dashed line) calculated for the integrated luminosity $L=10$~fb$^{-1}$
with the dipole form-factor. The unitarity curve (solid line) is calculated for $W_{\gamma\gamma} = 
\Lambda$, and due to inclusion of the from-factor is four times higher than in~Fig.~\ref{fig:a0w_noFF}.
The solid arrows indicate the strongest limit without violating the unitarity, and the corresponding energy scale $\Lambda$ 
of the new physics. Dashed arrow recalls the limit obtained without the form-factors. }
\label{fig:a0w_FF2TeV}
\end{figure}

\begin{figure}[htb]
\vspace{9pt}
  \includegraphics[width=19pc]{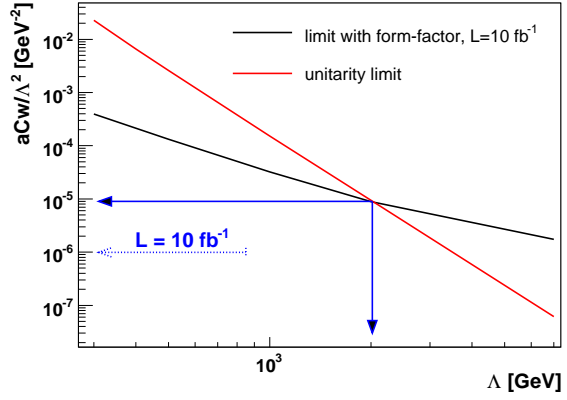}
\caption{The sensitivity plot for the W charged anomalous quartic coupling.
The plot is analogous to Fig.~\ref{fig:a0w_FF2TeV}.
}
\label{fig:aCw_FF2TeV}
\end{figure}

The sensitivity analysis has been repeated for the 
process $\gamma\gamma \rightarrow W^+W^-$ using the above form-factors. In 
Fig.~\ref{fig:a0w_FF2TeV} and in Fig.~\ref{fig:aCw_FF2TeV} the new sensitivity 
limits for the integrated luminosity $L=10$~fb$^{-1}$ are shown. 

The limits were calculated for 5 different $\Lambda$ values and linearly interpolated.
Moreover, the unitarity limit (solid line in Figs.~\ref{fig:a0w_FF2TeV},\ref{fig:aCw_FF2TeV})
is calculated at $W_{\gamma\gamma}=\Lambda$, after correcting for the form-factor suppression. 
This choice can be explained in the following way. If one obtains a limit of an anomalous 
coupling for $W_{\gamma\gamma}=\Lambda$ than for the energies below $\Lambda$ the unitarity is
automatically preserved. For higher energies, bigger than $\Lambda$ the form-factor suppression 
should be strong enough to keep the unitarity unbroken.

By repeating the sensitivity analysis from Sec. 3, the results
shown in Fig.~\ref{fig:a0w_FF2TeV} and in Fig.~\ref{fig:aCw_FF2TeV}
are obtained, and the following 95\% CL limits for the anomalous quartic couplings can be 
derived for $\gamma\gamma\rightarrow W^+W^-$ fully leptonic events, for 
the integrated luminosity $L=10$~fb$^{-1}$: 
\begin{eqnarray}
a_0^W/\Lambda^2 & < & 2.5\cdot 10^{-6}~\mathrm{GeV}^{-2}\\
a_C^W/\Lambda^2 & < & 9\cdot 10^{-6}~\mathrm{GeV}^{-2},
\end{eqnarray}
which are about 4~000 times stronger than achieved at LEP \cite{OPAL.limits},
and are ~10 times weaker than results from Tab.~\ref{tab.1st-limits}, obtained 
ignoring the unitarity constraints.
\section{Summary and perspectives}
The initial analysis of sensitivity to genuine AQGCs in photon-photon collisions at the LHC has
been presented. A simple method of event counting has provided very promising results, showing 
improvements of the LEP AQGC limits by a factor of about 4~000, for the integrated luminosity 10~fb$^{-1}$.
The effects of the form-factors, introduced to preserve unitarity constraints, are taken into account.
In addition, the analysis demonstrates the sensitivity to the new physics energy scale up to 2 TeV.

There are still several possible ways to improve the sensitivity. First, one can include also 
semi-leptonic channels in $\gamma\gamma\rightarrow WW$, which will increase statistics about 6 times.

\begin{figure}[htb]
\vspace{9pt}
  \includegraphics[width=19pc]{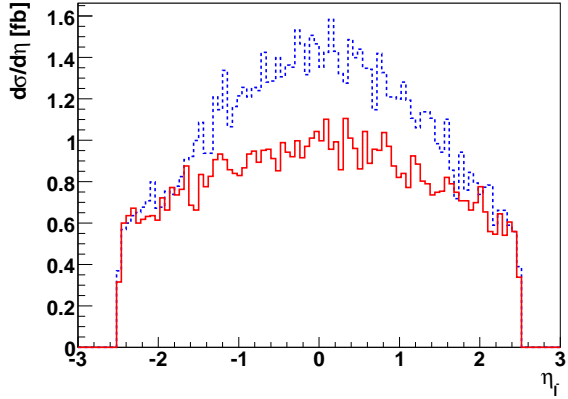}
\caption{Differential cross sections $d\sigma/d\eta$ for $\gamma\gamma\rightarrow W^+W^-\rightarrow l^+l^-$ after acceptance cuts
for the SM (solid line) and  for the anomalous quartic coupling (dashed line), $a_0^W/\Lambda^2=2.5\cdot 10^{-6}$~GeV$^{-2}$ 
and $\Lambda = 2$~TeV.}
\label{fig:Diff_eta}
\end{figure}

\begin{figure}[htb]
\vspace{9pt}
  \includegraphics[width=19pc]{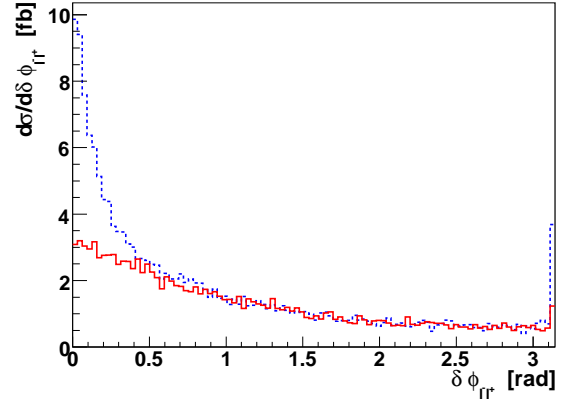}
\caption{Differential cross section $d\sigma/d(\delta\phi)$ for 
$\gamma\gamma\rightarrow W^+W^-\rightarrow  l^+l^-$ after acceptance cuts,
for the SM (solid line) and for the anomalous quartic coupling (dashed line),
$a_0^W/\Lambda^2=2.5\cdot 10^{-6}$~GeV$^{-2}$ with $\Lambda = 2$~TeV.}
\label{fig:Diff_acopla}
\end{figure}

In addition, one can increase sensitivity by studing the differential distributions, like the lepton 
pseudo-rapidity (see Fig.~\ref{fig:Diff_eta}), or the lepton acoplanarity $\delta \phi = 
\pi - Min(2\pi - \Delta \phi, \Delta \phi)$ (see Fig.~\ref{fig:Diff_acopla}) distributions. 
Finally, assuming the custodial SU(2)$_c$ symmetry, one can convert the limits
of $a_0^W$ and $a_0^Z$ into the single $a_0$ limit, and similarly the $a_C^W$ and $a_C^Z$ limits 
into the single $a_C$ limit (see. Eq. 1). This will be the subject of a future publication.

\end{document}